\documentclass[aps,preprint,showpacs,showkeys]{revtex4}
\usepackage{amsmath}
\usepackage{pstricks}
\usepackage{color}
\usepackage{graphicx}
\newcommand{\be}{\begin{equation}}
\newcommand{\ee}{\end{equation}}
\newcommand{\bea}{\begin{eqnarray}}
\newcommand{\eea}{\end{eqnarray}} 
\newcommand{\ba}{\begin{array}}
\newcommand{\ea}{\end{array}}

\newcommand{\bb}{\bibitem}
\begin{document}

\title{\bf Neumann boundary conditions with null external quasi-momenta in finite-systems} 
\author{Messias V. S. Santos\footnote{e-mail:messiasvilbert@df.ufpe.br}, Jos\'e B. da Silva Jr.\footnote{e-mail:jborba@petrobras.com.br}, Marcelo M. Leite\footnote{e-mail:mleite@df.ufpe.br}}
\affiliation{{\it Laborat\'orio de F\'\i sica Te\'orica e Computacional, Departamento de F\'\i sica,\\ Universidade Federal de Pernambuco,\\
50670-901, Recife, PE, Brazil}}
\vspace{0.2cm}
\begin{abstract}
{\it The order parameter of a critical system defined in a layered parallel plate geometry subject to Neumann boundary conditions at the limiting surfaces is studied. We utilize a one-particle irreducible vertex parts framework in order to study the critical behavior of such a system. The renormalized vertex parts are defined at zero external quasi-momenta, which makes the analysis particularly simple. The distance between the boundary plates $L$ characterizing the finite size system direction perpendicular to the hyperplanes plays a similar role here in comparison with our recent unified treatment for Neumann and Dirichlet boundary conditions. Critical exponents are computed using diagrammatic expansion at least up to two-loop order and are shown to be identical to those from the bulk theory (limit $L \rightarrow \infty$).}      
\end{abstract}
\vspace{1cm}
\vspace{1cm}

\maketitle

\newpage
\section{Introduction}
\par Finite size systems have a great deal of interest to our comprehension of several phenomena, particularly those involving critical systems \cite{Fisher,Barber,Privman}. The main question is: how far can one reduce the size of a system without disturbing its critical universal properties? Consider three-dimensional infinite (bulk) systems in a cubic lattice. If one 
dimension perpendicular to the planes of the critical material slab is reduced, one has a geometry of thin films, the critical temperature is lower in comparison with that from the bulk and the system still has divergences in its thermodynamic potentials. The thermodynamic limit can be safely taken in these thin films provoking the physical divergences, in stark contrast with thin films which are finite in all directions. In the latter, the thermodynamic 
limit {\it cannot be taken} and instead of the cusp-like divergences, the maximum corresponding to the critical temperature is no longer discontinous but it is "rounded" \cite{Gasparini}. We can also consider systems with two reduced dimensions and one infinite dimension. Their classical phase transitions effects appear, for instance in a cylindrical geometry regarding $Fe$ nanowires \cite{SU}, in the study of Potts model in long cylinders \cite{B}, etc. 
\par From the experimental point of view, the fabrication of high-quality thin films is commonly based on epitaxial growth over a substrate \cite{Falicov}. In general, ferromagnetic properties of epitaxially grown thin films depend upon the substrate. If the substrate forms a heterostruture with the original film, for instance, in $La_{2}NiMnO_{6}/SrTiO_{3}$  \cite{Jin}, ferrolectricity is induced in the heterostructure due to epitaxial strain in the $La_{2}NiMnO_{6}$ planes (which is originally bulk ferromagnetic). Depending on the method chosen and the substrate where the film is epitaxially grown, the critical properties of the bulk can be very different from that of the film itself: there can be effects of interfaces \cite{Monsen} (and its applications in the electroceramics industry \cite{Schlom}), surfaces \cite{Falicov,D}, etc. The parallel plate geometry is certainly important in the construction of microscopic devices as a stack of several layers of material in manufacturing, for example, high performance disks for memory storage \cite{HBBHBM}. In particular, multiferroic materials show ferromagnetism and ferroelectricity and are ideal for that purpose \cite{Scott}, such that the comprehension of those effects in finite-systems would be worthwhile. For a recent overview with a broader perspective and a mix of applications, see for instance \cite{subra}.
\par However, the epitaxial growth method of thin films named metal-organic chemical vapor deposition ($MOCVD$) is rather efficient to produce high-quality films with bulk properties, since the substrate can be "sheared" in the final process. This technique was employed in single crystal films of $Fe$ \cite{Kaplan} and $Ni$ \cite{Hong}. The post-growth analysis exhibited normal bulk magnetic properties. (There are also examples of materials whose molecular structure is composed by 
different atomic species in which this method of producing thin films show similar magnetic properties with those from bulk samples, albeit not so simply; see Ref. \cite{Ihzaz}). Hence, we learn that thin films produced in this manner yields a firm basis to investigate finite size effects realized experimentally in a simplified way. The field-theoretical description of critical properties for this geometry utilizing Green's functions in momentum space was initiated by Nemirovsky and Freed \cite{NF1,NF2} utilizing a massive method involving only bulk fields with several boundary conditions on the limiting surfaces. Recently, the renormalized one-particle irreducible ($1PI$) framework was devised and extended to massless fields as well, subject to periodic ($PBC$) and antiperiodic ($ABC$) boundary conditions \cite{BL}. After that, the $1PI$ vertex part formalism was generalized to include Dirichlet ($DBC$) and Neumann ($NBC$) boundary conditions in a non-trivial unified fashion by renormalizing the primitively divergent vertex parts at nonzero external quasi-momenta \cite{SBL}. The simplest situation within this picture is realized when each plate has linear dimensions of infinite extent. These thin films  satisfy the thermodynamic limit and were shown to have critical exponents identical to bulk systems and independent of boundary conditions, using massive or massless regimes for the renormalized volumetric order parameters (fields), provided the distance between the limiting surfaces is not too small. 
\par In this work we propose the renormalized one-particle ($1PI$) irreducible formalism at null external quasi-momenta in the treatment of a critical system defined in a parallel plate geometry, whose order parameter is subject to Neuman boundary conditions ($NBC$) at the limiting surfaces constituted by $(d-1)$-dimensional hyperplanes located at $z=0,L$, where $z$ is the space dimension perpendicular to the plates. We see finite size corrections to the bulk behavior which are $L$-dependent. In the limit $L \rightarrow \infty$ we show that the bulk behavior is recovered and all corrections tend to zero. Contrarily, in the limit $L \rightarrow 0$ we demonstrate how the "dimensional crossover"  \cite{Gasparini,Mockler1,Mockler2} shows up, overcoming the bulk behavior through non-trivial surface left-over, entirely out of bulk fields. Differently from previous approaches, the appearance of surface effects is due neither to the presence of interfaces between two materials (or the same material in different structural phases) nor the presence of external surface fields. Interestingly, surface effects emerge through dimensional crossover for $PBC$ and $NBC$ but are absent 
(or are much weaker, so to speak) in $DBC$ and $ABC$. 
\par In contrast with the situation for $PBC$ and $ABC$ which preserve translation invariance \cite{BL}, we will show that in the present framework $NBC$ do break the translation invariance of the theory in line with the result from the unified treatment \cite{SBL}. Moreover, the present method is by far simpler than the latter in the treatment of $NBC$. If the system is kept away from the dimensional crossover region, we prove that the critical exponents are the same of those from the bulk system, which is consistent with the results from the above previous $1PI$ vertex part and independent of boundary conditions. 
\par The ideas and results are presented as follows. We write down the bare Lagrangian density out of solely bulk fields 
and offer a quick review of the decomposition of the fields in terms of its Fourier components and basis functions in Section II. Section III will be the starting point with the basis functions from the Lagrangian density in the exponential representation. We write down the tensors corresponding to $1PI$ vertex parts (with and without composite operators). Utilizing the zero mode basis function as well, due to our choice of normalization conditions, we find later the complete set of Feynman graphs up to three-loop order. In Section IV we tie up the formulation of massive fields with a quick analysis of the one-loop diagram for the coupling constant and prove that the finite size corrections do not require normal ordering at zero external quasi-momenta. We show that the limit of validity of the $\epsilon$-expansion can be interpreted  consistently with that from the unified technique and compute the critical exponents. In Appendix A we summarize the results of all massive integrals. In Section V we introduce the massless framework in parallel with the massive case. Thereby, the critical exponents are shown to be identical in both massive and massless formulations. They agree exactly with those obtained from the infinite (bulk) system. The results of all massless diagrams are presented in Appendix B. The conclusions are the subject of Section VI.
\section{Preliminaries}
\par The bare Lagrangian density involving only volumetric (bulk) fields in the parallel plate geometry is given by:
\begin{equation}\label{1}
\mathcal{L} = \frac{1}{2}
|\bigtriangledown \phi_{0}|^{2} + \frac{1}{2} \mu_{0}^{2}\phi_0^{2} + \frac{1}{4!}\lambda_0(\phi_0^{2})^{2} .
\end{equation}
Here $\phi_{0}$, $\mu_{0}$ and $\lambda_{0}$ are the bare order parameter, mass 
($\mu_{0}^{2}= t_{0}$ is the bare reduced temperature proportional to $(\frac{T-T_{C}}{T_{C}})$) and coupling 
constant, respectively. Note that although $T_{C}$ denotes the critical temperature of the infinite system ("bulk") 
and is in general different of the shifted critical temperature of the finite system $T_{C}(L)$, in principle the difference is not sufficient to provoke a functional variation in the critical exponents at the critical region. Obviously, proving this with quantitative computation of the critical exponents is appropriate, but it is reasonable to note that the fluctuations there do not distinguish between these two types of temperatures \cite{NF1,NF2,JGH}, provided the plates are of infinite extent in each linear dimension. As discussed above, if the plates are not infinite, instead of a spike-like divergence of the critical quantity under consideration with its associated critical exponent, the finite size of the system provokes a "rounding" in the divergence. On aforementioned experimental grounds,  the temperature $T_{C}(L)$ yields a qualitative picture which resembles that from the bulk system. We shall use henceforth the bulk critical temperature in what follows. We refer the reader to Ref. \cite{SBL} in order to fix the notation, which is quite similar to that we shall employ hereafter.  
\par The field is a vector of $N$ components $((\phi_{0}^{2})^{2}= (\phi_{01}^{2} +...+\phi_{0N}^{2})^{2})$, whose internal indices of the $O(N)$ symmetry are omitted. The coordinate vector $\vec{\rho}$ belonging to the $(d-1)$-dimensional plates along with the perpendicular $z$ axis constitutes the collective vector $x=(\vec{\rho},z)$. There are parallel plates separated by a lattice constant space along the $z$ direction and filling in the region between  $z=0$ and $z=L$. The field satisfies $\frac{\partial \phi_{0}}{\partial z}(z=0)= \frac{\partial \phi_{0}}{\partial z}(z=L)=0$ 
for $NBC$. 
\par The order parameter is related to its Fourier modes in momentum space through 
$\phi_{0}(x)= {\underset{j}{\sum}} \int d^{d-1}k exp(i\vec{k}.\vec{\rho}) u_{j}(z) \phi_{0j}(\vec{k})$, where $\vec{k}$ is the momentum vector characterizing the $(d-1)$-dimensional space. The basis functions $u_{j}(z)$ specify the (hyper)plate and have a discrete index, with eigenvalues  $\kappa_{j}^{2}$, where 
$\kappa_{j}=\frac{\pi j}{L} \equiv {\tilde{\sigma} j}$ $(j=0,1,2...)$ is the quasi-momentum along the $z$-direction. The free bare massive ($\mu_{0}^{2} \neq 0$) propagator in momentum space is given by the expression 
$G_{0}(k,j, \mu_{0}) = \frac{1}{k^{2} + \tilde{\sigma}^{2}j^{2} 
+ \mu_{0}^{2}}$. The propagator can be represented graphically by a line with two extremities: the left one characterized by the index $i$ and the right labeled by $j$, which implies the presence of the overall factor $\delta_{ij}$, which 
will be ignored in what follows without loss of generality. The propagator for the massless case is obtained by setting 
$\mu_{0}^{2}= 0$.
\par From our construction of a generic Feynman diagram previously in the unified formalism, we just highlight that each momentum line (propagator) 
must be multiplied by $S_{j_{1}j_{2}} = \int_{0}^{L} dz u_{j_{1}}(z)u_{j_{2}}(z)$ and the $\phi^{4}$ vertices 
are multipled by the tensor $S_{j_{1}j_{2}j_{3}j_{4}}= \int_{0}^{L} 
dz u_{j_{1}}(z)u_{j_{2}}(z)u_{j_{3}}(z)u_{j_{4}}(z)$. Recall that this is the way in which the finite size effect is implemented as an internal symmetry, consisting of the direct product of these tensors with the usual ones from the 
$O(N)$ symmetry. 
\par The basis functions  for $NBC$ have the nonzero modes $u_{j}(z)= \Bigl(\frac{2}{L}\Bigr)^{\frac{1}{2}} cos(\kappa_{j} z)$ ($j=1,2...$) as well as the zero mode $u_{0}= \Bigl(\frac{1}{L}\Bigr)^{\frac{1}{2}}$. The explicit choice of vanishing external quasi-momenta in the renormalized theory implies that the zero mode basis function participates in the Feynman rules in a simpler but different form in comparison with the unified approach involving $DBC$ and $NBC$ renormalized at nonzero external quasi-momenta. 
\section{Feynman diagrams of the $1PI$ vertices $\Gamma^{(2)}, \Gamma^{(4)}$ and $\Gamma^{(2,1)}$}
\par  Let us start the construction representing mathematically the several Feynamn diagrams required to our computation by writing down the tensors $S_{ij}$, $S_{ijkl}$ and $\hat{S}_{ijk}$ corresponding to the composite field. We multiply 
the tensors with four indices by $2 \pi$ in order to construct the diagrams, namely $\tilde{S}_{ijkl}= 2 \pi S_{ijkl}$. We are using explicitly the exponential representation for the nonzero mode basis functions; compare the tensors with those from Ref. \cite{SBL} when $\tau=1$. The explicit form of the tensors required using the above eigenfunctions are 
$(\delta_{i+j+k+l,0} \equiv \delta(i+j+k+l), \tilde{\sigma}=\frac{\pi}{L})$:
\begin{subequations}
\begin{eqnarray}\label{2}
&& S_{i_{1} i_{2}}= \delta(i_{1} - i_{2}) + \delta(i_{1} + i_{2}), \;\; S_{00} = 1\\
&& \tilde{S}_{i_{1} i_{2} i_{3} i_{4}}= \tilde{\sigma} [\delta(i_{1} + i_{2} + i_{3} + i_{4}) + \delta(i_{1} - i_{2} + i_{3} + i_{4}) + \delta(i_{1} + i_{2} - i_{3} + i_{4}) + \delta(i_{1} + i_{2} \nonumber\\
&& \;\; + \;\; i_{3} - i_{4}) + \delta(i_{1} - i_{2} - i_{3} + i_{4}) + \delta(i_{1} - i_{2} + i_{3} - i_{4}) + \delta(i_{1} + i_{2} - i_{3} - i_{4}) \nonumber\\
&& \;\; + \;\; \delta(i_{1} - i_{2} - i_{3} - i_{4})],\\
&& \tilde{S}_{0 i_{1} i_{2} i_{3}}= \sqrt{2} \tilde{\sigma} [\delta(i_{1} + i_{2} + i_{3}) + \delta(i_{1} + i_{2} - i_{3}) + \delta(i_{1} - i_{2} + i_{3}) + \delta(i_{1} - i_{2} - i_{3})],\\
&& \tilde{S}_{0 0 i_{1} i_{2}} = 2 \tilde{\sigma} [\delta(i_{1} - i_{2}) + \delta(i_{1} + i_{2})], \;\;\tilde{S}_{0 0 0 i_{1}} = 0,\;\; \tilde{S}_{0 0 0 0} = 2 \tilde{\sigma},\\
&& \hat{S}_{i_{1} i_{2} i_{3}}=\frac{1}{2}[\delta(i_{1} + i_{2} + i_{3}) + \delta(i_{1} + i_{2} - i_{3}) + \delta(i_{1} - i_{2} + i_{3}) + \delta(i_{1} - i_{2} - i_{3})],\\
&& \hat{S}_{0 i_{1} i_{2}}= \frac{1}{\sqrt{2}}[\delta(i_{1} - i_{2}) + \delta(i_{1} + i_{2})], \;\; \hat{S}_{00i}= \delta(i). 
\end{eqnarray}
\end{subequations}
Note that we separated the zero mode in all tensors above (the other indices appearing in the above expressions 
are explicitly nonzero).
\par There is an argument presented in the unified picture which simplifies enormously our task in evaluating Feynman graphs out of these tensors. We shall outline it very briefly here and take advantage of it in our present discussion. In order to restrict our discussion to a minimal number of diagrams we start with a tree level bare mass parameter $\mu_{0}$, perform the diagrammatic expansion of the $\Gamma^{(2)}(\vec{k}=0;i=0; \tilde{\sigma}, \mu_{0})$ and identify it with the three-loop bare mass $\mu$. After the inversion in order to obtain $\mu_{0} = \mu_{0}(\mu)$ and expressing all integrals in terms of $\mu$, this reparameterization in conjunction with the non-trivial tadpole cancelation eliminates all tadpole diagrams and tadpole mass insertions in all primitively divergent diagrams, {\it i. e.}, $\Gamma^{(2)}, \Gamma^{(4)}$ and $\Gamma^{(2,1)}$ \cite{SBL,CL}. Therefore, we just have to analyze the "sunset" (two-loop) and the nontrivial three-loop diagram of  $\Gamma^{(2)}(\vec{k};i=0; \tilde{\sigma},\mu)$. In addition, we are left only with the evaluation of the diagrams from $\Gamma^{(4)}(\vec{k}_{i}; 0,0,0,0; \tilde{\sigma}, \mu)$ $(i=1,...,4)$ and $\Gamma^{(2,1)}(k_{1},k_{2}, Q;0,0,0; \tilde{\sigma}, \mu)$ up to two-loop order without any mass insertions.
\par Before analyzing any loop diagram let us determine the value of all tree-level primitively divergent bare vertex parts at zero external quasi-momenta. First, the bare tree diagram of $\Gamma^{(2)}(\vec{k};i=0; \tilde{\sigma},\mu)$ is just equal to the inverse propagator, i.e., $k^{2} + \mu^{2}$ and represented by the graph
$ \parbox{10mm}{\includegraphics[scale=0.75]{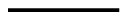}}^{-1}$. At this point is convenient switching to a simpler 
notation for the momenta $\vec{k} \equiv k$. The bare vertex 
$\Gamma^{(4)}(k_{i};0,0,0,0;\tilde{\sigma},\mu)$ starts with the zero-loop contribution $\parbox{5mm}{\includegraphics[scale=0.05]{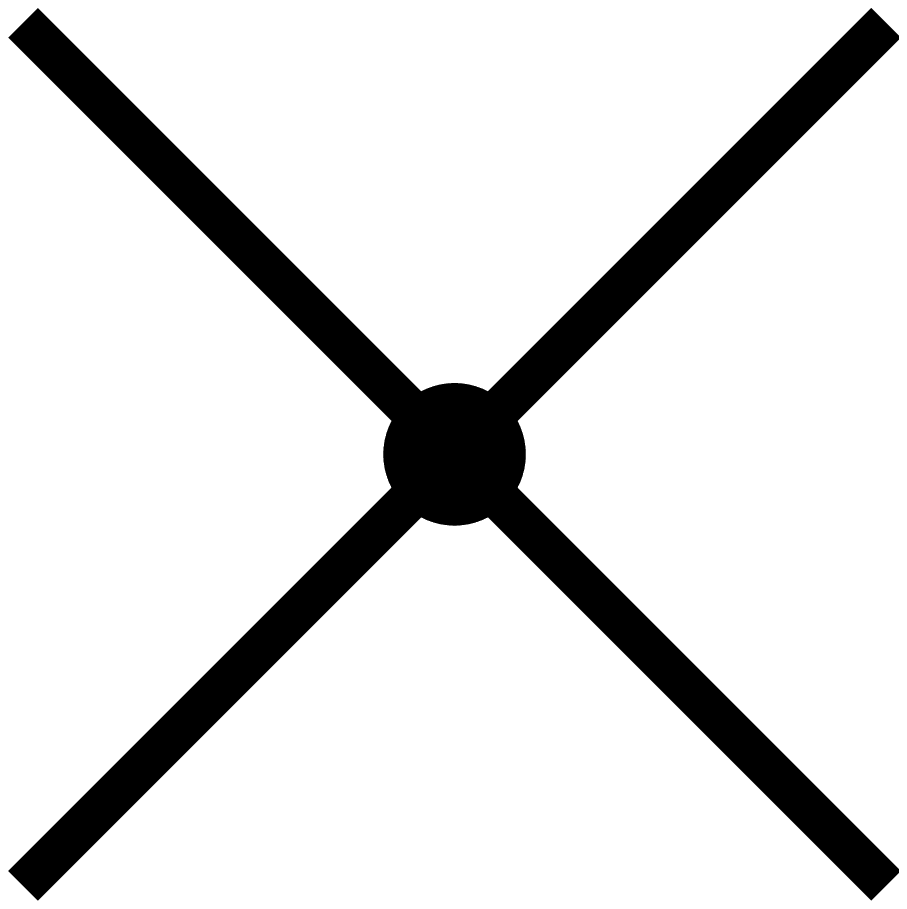}}=2 \lambda \tilde{\sigma}$, whereas the composite vertex operator $\Gamma^{(2,1)}(k_{1},k_{2}, Q; 0,0,0; \tilde{\sigma},\mu)$ has tree-level value 
$\parbox{5mm}{\includegraphics[scale=0.75]{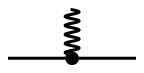}}\qquad=1$ \cite{Messias}. Therefore, differently from the unified approach, here each $l$-loop diagram from $\Gamma^{(4)}(k_{i};0,0,0,0;\tilde{\sigma},\mu)$ receives a coefficient $ 2 \lambda \tilde{\sigma}$, such that  this coefficient factors out in the diagrammatic expansion of this vertex part. We shall omit this trivial global factor in the construction of arbitrary loop diagrams from $\Gamma^{(4)}(k_{i};0,0,0,0;\tilde{\sigma},\mu)$, but will recuperate it during the discussion of the normalization conditions later on. We also choose not to write explicitly the coupling constant in front of each diagram; rather we use their different powers as coefficients of the graphs (see below). The other vertex parts $\Gamma^{(2)}(k;i=0; \tilde{\sigma},\mu)$ and $\Gamma^{(2,1)}(k_{1},k_{2},Q;0,0,0; \tilde{\sigma},\mu)$ do not receive global factors in arbitrary loop order and have pretty much the same form as in the usual $\phi^{4}$ theory describing bulk systems. Although not depicted in the diagrams, we will employ the convention that all external quasi-momenta of all diagrams (obviously including those from tree level contributions) are implicitly set to zero.       
\par We analyze the minimal set of diagrams from $\Gamma^{(2)}(k;i=0; \tilde{\sigma}, \mu)$ to begin with. The preliminary expression for its two-loop diagram  is given by:
\begin{eqnarray}\label{3}
&& \parbox{10mm}{\includegraphics[scale=1.2]{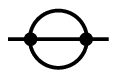}} \quad = \Bigl(\frac{N+2}{3}\Bigr) \overset{\infty}{\underset{j_{1},j_{2},j_{3}=0}{\sum}} \tilde{S}_{0 j_{1} j_{2} j_{3}} \tilde{S}_{j_{1} j_{2} j_{3} 0} \int d^{d-1}q_{1} d^{d-1}q_{2} G_{0}(q_{1}+q_{2}+k, j_{3}) G_{0}(q_{2}, j_{2}) \nonumber\\
&& \qquad \times \;\; G_{0}(q_{1}, j_{1}). 
\end{eqnarray}
\par After explicitly separating the zero mode components from the nonzero ones and perform some manipulations one can show that this expresssion reduces to
\begin{eqnarray}\label{4}
&& \parbox{10mm}{\includegraphics[scale=1.2]{fig6.eps}}  \quad = \left(\frac{N+2}{3}\right)\Bigl[I_{3}(k;0;\tilde{\sigma},\mu) + 3 \tilde{I}_{3}(k;0,0;\tilde{\sigma},\mu)\Bigr]
\end{eqnarray}
where,
\begin{subequations}\label{5}
\begin{eqnarray}
&& I_{3}(k;0;\tilde{\sigma},\mu) = \tilde{\sigma}^{2} \overset{\infty}{\underset{j_{1},j_{2}=-\infty}{\sum}} 
\int \frac{d^{d-1}q_{1} d^{d-1}q_{2}}
{[q_{2}^{2}+\tilde{\sigma}^{2}j_{2}^{2}+\mu^{2}][(q_{1}+q_{2}+k)^{2}
+\tilde{\sigma}^{2}(j_{1}+j_{2})^{2}+\mu^{2}]}\nonumber\\
&& \;\;\;\quad \times \frac{1}{[q_{1}^{2}+\tilde{\sigma}^{2}j_{1}^{2}+\mu^{2}]},\\
&& \tilde{I}_{3}(k;0,0;\tilde{\sigma},\mu)= \tilde{\sigma}^{2} \overset{\infty}{\underset{l=-\infty}{\sum}} 
\int \frac{d^{d-1}q_{1} d^{d-1}q_{2}}
{[q_{1}^{2}+\tilde{\sigma}^{2}l^{2}+\mu^{2}][q_{2}^{2}+\mu^{2}]} \frac{1}{[(q_{1}+q_{2}+k)^{2}
+\tilde{\sigma}^{2} l^{2}+\mu^{2}]}.
\end{eqnarray}
\end{subequations}
\par Interestingly, the first integral is just identical to the similar contribution coming from $PBC$ at zero external quasi-momenta \cite{BL}, whereas the second one correspond to the "non-diagonal" term which breaks explicitly the translational invariance at zero external quasi-momenta. The latter just shows up for $NBC$ and $DBC$ and was discussed 
previously in the unified picture with nonvanishing external quasi-momenta. However, the present form is much simpler 
than in the situation when the external quasi-momenta are not chosen equal to zero.
\par Let us analyze now the three-loop contribution of this vertex part at zero external quasi-momenta. In terms of the finite size tensors, it reads
\begin{eqnarray}\label{6}
&& \parbox{10mm}{\includegraphics[scale=1.2]{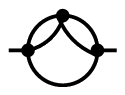}}  \quad =  \frac{(N+2)(N+8)}{27} \overset{\infty}{\underset{j_{1},j_{2},j_{3}, j_{4}, j_{5}=0}{\sum}} \tilde{S}_{0 j_{1} j_{2} j_{3}} \tilde{S}_{j_{2} j_{3} j_{4} j_{5}} \tilde{S}_{j_{4} j_{5} j_{1} 0} \int d^{d-1}q_{1} d^{d-1}q_{2} d^{d-1}q_{3} G_{0}(q_{1}, j_{1}) \nonumber\\
&& \qquad \qquad \times \;\; G_{0}(q_{1}+q_{2}+k, j_{2}) G_{0}(q_{2}, j_{3}) G_{0}(q_{1}+q_{3}+k, j_{4}) G_{0}(q_{3}, j_{5}).
\end{eqnarray}
\par The manipulation here involving the separation of zero modes and nonzero modes components of each index in the summation of the finite size tensors is really lengthy and we let the details to the reader figure them out. Fortunately, after a formidable large number of cancelations, the outcome turns out to be simple and we get to 
\begin{eqnarray}\label{7}
&& \parbox{10mm}{\includegraphics[scale=1.2]{fig7.eps}}  \quad =  \frac{(N+2)(N+8)}{27}
[I_{5}(k,0,\tilde{\sigma},\mu) + \tilde{I}_{5}(k,0,0,\tilde{\sigma},\mu) + 4 \hat{I}_{5}(k,0,0,0,\tilde{\sigma},\mu)\nonumber\\
&& \;\;+\;\; 2 \bar{I}_{5}(k,0,0,0,\tilde{\sigma},\mu)].
\end{eqnarray}
The integrals in last equation are defined by:
\begin{subequations}\label{8}
\begin{eqnarray}
&& I_{5}(k;0;\tilde{\sigma},\mu)= \tilde{\sigma}^{3} \overset{\infty}{\underset{j_{1},j_{2},j_{3}=-\infty}{\sum}} 
\int \frac{d^{d-1}q_{1} d^{d-1}q_{2}d^{d-1}q_{3}}
{[q_{1}^{2}+\tilde{\sigma}^{2}j_{1}^{2}+\mu^{2}][q_{2}^{2}+\tilde{\sigma}^{2}j_{2}^{2}+\mu^{2}][q_{3}^{2}+\tilde{\sigma}^{2}j_{3}^{2}+\mu^{2}]}\nonumber\\
&& \times \frac{1}{[(q_{1}+q_{2}+k)^{2}
+\tilde{\sigma}^{2}(j_{1}+j_{2})^{2}+\mu^{2}][(q_{1}+q_{3}+k)^{2}
+\tilde{\sigma}^{2}(j_{1}+j_{3})^{2}+\mu^{2}]},\\
&& \tilde{I}_{5}(k;0,0;\tilde{\sigma},\mu)= \tilde{\sigma}^{3} \overset{\infty}{\underset{j_{1},j_{2}=-\infty}{\sum}} 
\int \frac{d^{d-1}q_{1} d^{d-1}q_{2}d^{d-1}q_{3}}
{[q_{1}^{2}+\mu^{2}][q_{2}^{2}+\tilde{\sigma}^{2}j_{1}^{2}+\mu^{2}][q_{3}^{2}+\tilde{\sigma}^{2}j_{2}^{2}+\mu^{2}]}\nonumber\\
&& \times \frac{1}{[(q_{1}+q_{2}+k)^{2}
+\tilde{\sigma}^{2} j_{1}^{2}+\mu^{2}][(q_{1}+q_{3}+k)^{2}
+\tilde{\sigma}^{2} j_{2}^{2}+\mu^{2}]},\\
&& \hat{I}_{5}(k;0,0,0; \tilde{\sigma}, \mu)=\tilde{\sigma} \overset{\infty}{\underset{j=-\infty}{\sum}} 
\int \frac{d^{d-1}q \tilde{I}_{2}(q+k;0,j;\tilde{\sigma}, \mu) I_{2}(q+k;j;\tilde{\sigma}, \mu)}{q^{2} + \sigma^{2}j^{2} + \mu^{2}},\\
&& \bar{I}_{5}(k;0,0,0;\tilde{\sigma},\mu)= \tilde{\sigma} \overset{\infty}{\underset{j_{1},j_{2}=-\infty}{\sum}} 
\int \frac{d^{d-1}q \tilde{I}_{2}(q+k;j_{2};j_{2}+j_{1}; \tilde{\sigma}, \mu)}{q^{2} + \sigma^{2}j_{1}^{2} + \mu^{2}} \nonumber\\
&& \;\; \times \;\; \tilde{I}_{2}(q+k;j_{1}+j_{2},j_{2}; \tilde{\sigma}, \mu).
\end{eqnarray}
\end{subequations}
\par It is important to emphasize that the integrals $I_{2}(k;0;\tilde{\sigma},\mu)$ and 
$\tilde{I}_{2}(k;0,0;\tilde{\sigma},\mu)$ correspond to one-loop subdiagrams of the vertex part $\Gamma^{(4)}(k_{i};0,0,0,0; \tilde{\sigma}, \mu)$ and are defined by
\begin{subequations}
\begin{eqnarray}\label{9}
&& I_{2}(k;0;\tilde{\sigma},\mu) = \tilde{\sigma} \overset{\infty}{\underset{j=-\infty}{\sum}} 
\int \frac{d^{d-1}q}{[(q+k)^{2} + \tilde{\sigma}^{2} j^{2} + \mu^{2}][q^{2} + \tilde{\sigma}^{2} j^{2} + \mu^{2}]},\label{9a}\\
&& \tilde{I}_{2}(k;0,0;\tilde{\sigma},\mu) = \tilde{\sigma} 
\int \frac{d^{d-1}q}{[(q+k)^{2} + \mu^{2}][q^{2} + \mu^{2}]}.\label{9b}
\end{eqnarray}
\end{subequations}
The pattern of $PBC$ (integral $I_{5}$) plus "non-diagonal" terms happens in a way consistent with the unified framework 
at nonzero external quasi-momenta for $NBC$. That is why we included an extra entry of quasi-momenta in the the integral 
$\tilde{I}_{2}(k;0,0;\tilde{\sigma},\mu)$. This is a generic feature of the non-diagonal terms showing up in arbitrary loop diagrams: if the diagonal contribution (integral) has $n$ entries in its argument for the external quasi-momenta, the nondiagonal contribution will have $n+1$ entries. This fact is trivial when addressing all the contributions at zero external momenta, but should be kept in mind in order to make contact with the more general unified formalism.
\par We now discuss the vertex part $\Gamma^{(4)}(k_{i};0,0,0,0; \tilde{\sigma}, \mu)$. The one-loop diagram in terms of 
the finite size tensors and propagators can be written as  
\begin{eqnarray}\label{10}
&& \parbox{10mm}{\includegraphics[scale=1.00]{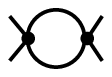}}\quad = \frac{(N+8)}{9} 
\overset{\infty}{\underset{l_{1},l_{2} = 0}{\sum}}\tilde{S}_{00 l_{1} l_{2}} \tilde{S}_{l_{1} l_{2}00} \int d^{d-1}q G_{0}(q+k,l_{1}) G_{0}(q,l_{2}).
\end{eqnarray}  
We employ the standard notation that in this diagram the external momentum $k$ can be connected with the actual external momenta $k_{i}, (i=1,2,3,4)$ through the possibilities $k=k_{1}+k_{2}$ , $k=k_{1}+k_{3}$ and $k=k_{2}+k_{3}$. (We could also have used $k_{4}$ due to the overall external momenta conservation.) All of them should be included in the computation of the contribution of this diagram. The same happens in the trivial two-loop diagram to be discussed below. However, for the two-loop non-trivial diagram of this vertex part, there are twice more terms in the complete computation due to these diagrams in the diagrammatic expansion. For the sake of simplicity we adopt the notation $k$ to designate a generic external momenta (and adopt a similar convention for the diagrams involving composite fields; see below). Since we are going to be interested in particular configurations of the external momenta ("symmetry points") in the renormalization scheme, we shall not worry about these details. The reader is invited to consult the book by Amit and Martin-Mayor \cite{Amit} at this point.  
\par We can handle this expression in order to get the expressions 
\begin{eqnarray}\label{11}
&& \parbox{10mm}{\includegraphics[scale=1.00]{fig10.eps}}\quad = \frac{(N+8)}{9} 
\Bigl(I_{2}(k;0; \tilde{\sigma}, \mu) + \tilde{I}_{2}(k;0,0;\tilde{\sigma}, \mu)\Bigr),
\end{eqnarray} 
where these integrals were defined in Eqs. (\ref{9a}) and (\ref{9b}).
\par By the same token, the trivial two-loop diagram computed at zero external quasi-momenta can be written as 
\begin{eqnarray}\label{12}
&& \parbox{10mm}{\includegraphics[scale=1.2]{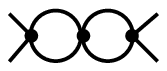}}\qquad = \frac{(N^{2}+6N+20)}{27} 
\overset{\infty}{\underset{j_{1},j_{2}, j_{3}, j_{4} = 0}{\sum}} \tilde{S}_{00 j_{1} j_{2}} \tilde{S}_{j_{1} j_{2} j_{3} j_{4}}\tilde{S}_{j_{3} j_{4} 00} \int d^{d-1}q_{1} d^{d-1}q_{2} G_{0}(q_{1}+k,j_{1}) \nonumber\\
&& \;\; G_{0}(q_{1},j_{2}) G_{0}(q_{2}+k,j_{3}) G_{0}(q_{1},j_{4}).
\end{eqnarray}   
\par When the same systematics is carried out by replacing the values of the tensors we are led to 
\begin{eqnarray}\label{13}
&& \parbox{10mm}{\includegraphics[scale=1.2]{fig11.eps}}\qquad = \frac{(N^{2}+6N+20)}{27}\Bigl[ I_{2}^{2}(k;0;\tilde{\sigma}, \mu) + 2 I_{2}(k;0;\tilde{\sigma}, \mu) \tilde{I}_{2}(k;0,0;\tilde{\sigma}, \mu) \nonumber\\
&& \;\; +\;\;  \overset{\infty}{\underset{j=-\infty}{\sum}}\tilde{I}_{2}^{2}(k;j,j;\tilde{\sigma}, \mu)\Bigr].
\end{eqnarray}
\par In order to complete our task of computing the two-loop contributions of the four-point vertex part, consider the non-trivial two-loop diagram 
\begin{eqnarray}\label{14}
&& \parbox{10mm}{\includegraphics[scale=1.2]{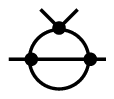}}\qquad = \frac{(5N+22)}{27} 
\overset{\infty}{\underset{j_{1},j_{2}, j_{3}, j_{4} = 0}{\sum}} \tilde{S}_{00 j_{1} j_{2}} \tilde{S}_{j_{1} j_{3} j_{4} 0}\tilde{S}_{j_{2} j_{3} j_{4} 0} \int d^{d-1}q_{1} d^{d-1}q_{2} G_{0}(q_{1}-k,j_{1}) \nonumber\\
&& \;\; G_{0}(q_{1},j_{2}) G_{0}(q_{1}-q_{2}+k_{3},j_{3}) G_{0}(q_{2},j_{4}).
\end{eqnarray}
\par After working out some set of calculations we find the simple result
\begin{eqnarray}\label{15}
&& \parbox{10mm}{\includegraphics[scale=1.2]{fig12.eps}}\qquad = \frac{(5N+22)}{27} 
\Bigl[I_{4}(k,k_{3};0,0;\tilde{\sigma},\mu) + \tilde{I}_{4}(k,k_{3};0,0,0;\tilde{\sigma},\mu) \nonumber\\
&& \;\;+\;\; 2 \hat{I}_{4}(k,k_{3};0,0,0;\tilde{\sigma},\mu)\Bigr],
\end{eqnarray}
where $k=k_{1}+k_{2}$ and the above integrals read
\begin{subequations}\label{16}
\begin{eqnarray}
&& I_{4}(k,k';0,0; \tilde{\sigma}, \mu)= \tilde{\sigma}^{2}\overset{\infty}{\underset{l,m=-\infty}{\sum}}\int \frac{d^{d-1}q_{1}d^{d-1}q_{2}}
{[q_{1}^{2}+\tilde{\sigma}^{2}l^{2}+\mu^{2}][(q_{1}-k)^{2}+\tilde{\sigma}^{2} l^{2}+\mu^{2}]
[q_{2}^{2}+\tilde{\sigma}^{2}m^{2}+\mu^{2}]}\nonumber\\
&& \;\;\times\;\; \frac{1}{[(q_{1}-q_{2}+k')^{2}+\tilde{\sigma}^{2}(l-m)^{2}+\mu^{2}]},\\
&& \tilde{I}_{4}(k,k';0,0,0; \tilde{\sigma}, \mu)=\tilde{\sigma}^{2}\overset{\infty}{\underset{m=-\infty}{\sum}}\int \frac{d^{d-1}q_{1}d^{d-1}q_{2}}
{[q_{1}^{2}+\mu^{2}][(q_{1}-k)^{2}+\mu^{2}]
[q_{2}^{2}+\tilde{\sigma}^{2}m^{2}+\mu^{2}]}\nonumber\\
&& \;\;\times\;\; \frac{1}{[(q_{1}-q_{2}+k')^{2}+\tilde{\sigma}^{2}m^{2}+\mu^{2}]},\\
&& \hat{I}_{4}(k,k';0,0,0; \tilde{\sigma}, \mu)=\tilde{\sigma}^{2}\overset{\infty}{\underset{m=-\infty}{\sum}}\int \frac{d^{d-1}q_{1}d^{d-1}q_{2}}
{[q_{1}^{2}+\tilde{\sigma}^{2}m^{2}+\mu^{2}][(q_{1}-k)^{2}+\tilde{\sigma}^{2}m^{2}+\mu^{2}]
[q_{2}^{2}+\mu^{2}]}\nonumber\\
&& \;\;\times\;\; \frac{1}{[(q_{1}-q_{2}+k')^{2}+\tilde{\sigma}^{2}m^{2}+\mu^{2}]}.
\end{eqnarray}
\end{subequations}
\par It is worthy noting that all integrals above are straightforward particular cases from their counterparts in the unified framework with nonvanishing external quasi-momenta \cite{SBL}. From this viewpoint, it is really rewarding to employ the present formalism to study $NBC$.
\par Let us examine the diagrams of the composite field. They have the same integral structure of the four-point integral just discussed. It will simplify the subsequent description, for we restrict ourselves to the explanation of the results in terms of the above integrals.  
\par We start with the tensors $\hat{S}_{ijk}$ whose contracted products produce the diagram with null external quasi-momenta. The one-loop graph
\begin{eqnarray}\label{17}
&& \parbox{13mm}{\includegraphics[scale=1.0]{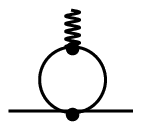}}\qquad = \frac{(N+2)}{6} \overset{\infty}{\underset{j_{1},j_{2} = 0}{\sum}}\tilde{S}_{00 j_{1} j_2} \hat{S}_{j_{1} j_{2} j} \int d^{d-1}q G_{0}(q+k,j_{1}) G_{0}(q,j_{2}),
\end{eqnarray}   
can be easily shown to be given by
\begin{eqnarray}\label{18}
&& \parbox{13mm}{\includegraphics[scale=1.0]{fig14.eps}}\qquad = \frac{(N+2)}{6}
\Bigl[I_{2}(k;0; \tilde{\sigma}, \mu) + \tilde{I}_{2}(k;0,0;\tilde{\sigma}, \mu)\Bigr],
\end{eqnarray} 
\par In order to make a connection with the unified framework, we treat both two-loop contributions for the composite field simultaneously. Their expressions with respect to the finite size tensors are
\begin{subequations}
\begin{eqnarray}\label{19}
&& \parbox{13mm}{\includegraphics[scale=1.0]{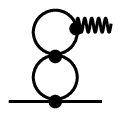}}\qquad = \frac{(N+2)^{2}}{108} \overset{\infty}{\underset{j_{1},j_{2}, j_{3}, j_{4} = 0}{\sum}} \tilde{S}_{00 j_{1} j_{2}} \tilde{S}_{j_{1} j_{2} j_{3} j_{4}} 
\hat{S}_{j_{3} j_{4} j} \int d^{d-1}q_{1} d^{d-1}q_{2} G_{0}(q_{1}+k,j_{1}) \nonumber\\
&& G_{0}(q_{1},j_{2})G_{0}(q_{2}+k,j_{3}) G_{0}(q_{2},j_{4}),\\
&& \parbox{13mm}{\includegraphics[scale=1.0]{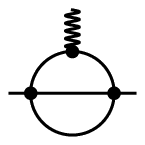}}\qquad = \frac{(N+2)}{36} \overset{\infty}{\underset{j_{1},j_{2}, j_{3}, j_{4} = 0}{\sum}} \tilde{S}_{0 j_{1} j_{3} j_{4}} \tilde{S}_{0 j_{2} j_{3} j_{4}} 
\hat{S}_{j_{1} j_{2} j} \int d^{d-1}q_{1} d^{d-1}q_{2} G_{0}(q_{1}-k,j_{1}) \nonumber\\
&& G_{0}(q_{1},j_{2})G_{0}(q_{1}-q_{2}+k_{3},j_{3}) G_{0}(q_{2},j_{4}).
\end{eqnarray} 
\end{subequations}
\par Working out the details encoded in the finite size tensors, the reader can check that
\begin{subequations}\label{20}
\begin{eqnarray}
&& \parbox{10mm}{\includegraphics[scale=1.0]{fig16.eps}} \qquad = \frac{(N+2)^{2}}{108} \Bigl[I_{2}^{2}(k;0; \tilde{\sigma}, \mu) + 2 I_{2}(k;0; \tilde{\sigma}, \mu) \tilde{I}_{2}(k;0,0; \tilde{\sigma}, \mu) ) \nonumber\\
&& + \overset{\infty}{\underset{l=-\infty}{\sum}} \tilde{I}_{2}^{2}(k;l,l; \tilde{\sigma}, \mu)\Bigr] ,\\
&& \parbox{10mm}{\includegraphics[scale=1.0]{fig17.eps}} \qquad = \frac{(N+2)}{36} \Bigl\{I_{4}(k,k_{3};0,0; \tilde{\sigma}, \mu) + \tilde{I}_{4}(k, k_{3};0,0,0;\tilde{\sigma},\mu) \nonumber\\
&& + 2 \hat{I}_{4}(k,k_{3};0,0,0; \tilde{\sigma}, \mu) \Bigr\}.
\end{eqnarray}
\end{subequations}
\par This concludes our discussion to getting all integrals required in the computation of critical exponents. The main difference with respect with other momentum space approaches for $NBC$ are the multiplicities of some $1PI$ primitively divergent vertex parts \cite{SBL,Messias}. We turn now our attention to the renormalization of the vertex parts in the massive and massless regimes. Our succint exposition of these topics herein should be complemented with the detailed account given in the unified approach involving $NBC$ and $DBC$.   
\section{Renormalization in the massive theory}
\subsection{Region of validity of the $\epsilon$-expansion} 
\par Let us focus on the integrals belonging to the 
one-loop diagram of the four-point function at zero external momenta. They will determine under what conditions the perturbative expansion is well-defined. Generically, two types of potential divergences are expected: the traditional ones realized as dimensional poles in $\epsilon=4-d$ and those directly connected with the lenght $L$. The latter appear as finite size corrections. Taking certain limits to $L$ the corrections either are well-behaved, thus validating perturbation theory, or their contribution in the integral blow up leading to a breakdown of the $\epsilon$-expansion. Their solutions are given in Appendix A. The combination appearing in the diagram yields:
\begin{subequations}\label{21}
\begin{eqnarray}
&& I_{2}(k=0;0; \tilde{\sigma}, \mu) + \tilde{I}_{2}(k=0;0;0; \tilde{\sigma}, \mu) =  \frac{\mu^{-\epsilon}}{\epsilon} \Bigl[1 - \frac{\epsilon}{2} + \epsilon \tilde{\zeta}_{0}(\tilde{r}) \Bigr], \label{21a}\\
&& \tilde{\zeta}_{0}(\tilde{r}) = \frac{1}{2} \Bigl[f_{\frac{1}{2}} (0, \tilde{r}^{-1}) + \tilde{r}\Bigr], \label{21b}\\
&& \tilde{r}=\frac{\tilde{\sigma}}{\mu},\label{21c}\\
&& f_{\alpha}(a,b)= 4\overset{\infty}{\underset{m=1}{\sum}} cos(2\pi ma) \bigl(\frac{\pi m}{b}\bigr)^{\alpha -\frac{1}{2}}K_{\alpha -\frac{1}{2}}(2 \pi mb),\label{21d}
\end{eqnarray}
\end{subequations}
where $K_{\nu}(x)$ is the modified Bessel function of second kind. Here $\tilde{\zeta}_{0}(\tilde{r})$ corresponds to the finite size correction. 
\par The limit $L \rightarrow \infty$ of this function was already studied in \cite{BL} in the cases of $PBC$ and $NBC$. Since $\tilde{r} \rightarrow 0$ in this limit, we collect together both outcomes to determine that  $\tilde{\zeta}_{0}(\tilde{r}) (L \rightarrow \infty) \rightarrow 0$ and correctly reduces to the bulk result.
\par Consider the $L \rightarrow 0$ limit. The functional form of the finite size correction is exceedingly simpler here: at zero external quasi-momenta we do not need to devise any "normal ordering" and we obtain
\begin{equation}\label{22}
\underset{L \rightarrow 0}{lim} \;\;  \tilde{\zeta}_{0}(\tilde{r})  \rightarrow \; ln \tilde{r}^{-1} +  \tilde{r},
\end{equation}
which is divergent, since $\tilde{r}(L \rightarrow 0) \rightarrow \infty$, therefore invalidating $\epsilon$-expansion perturbative results. The dominant divergence is that linear in $\tilde{r}$ and is responsible for the appearance of surface effects. The origin of this term comes not only from the translation invariant symmetry breaking piece from the integral $\tilde{I}_{2}$, but also from the "diagonal" integral $I_{2}$ \cite{BL}. They have the same sign and add up to produce this behavior. For the sake of comparison, for $PBC$ the coefficient of the linear term in $\tilde{r}(L \rightarrow 0) \rightarrow \infty$ is half the value of that from $NBC$, whereas in $DBC$ the diagonal contribution linear in $\tilde{r}$ coming from $I_{2}$ is exactly canceled by the contribution of the non-diagonal translational symmetry breaking integral $\tilde{I}_{2}$. That is why $DBC$ has the same logarithmic behavior in the limit $\tilde{r} \rightarrow \infty$ as in $ABC$. From this simple analysis we see the reason why $NBC$ rules the bulk-surface transition: the coefficient of the linear divergence when $L \rightarrow 0$ is twice as bigger compared with the one from $PBC$, preventing the latter to take over in the aforementioned structural phase transition.
\par Nevertheless, this behavior is in exact agreement with the unified approach for $NBC$ when normal ordering of the finite size correction is taken into account. This procedure had nothing to do with physical reasons. As pointed out before, had we treated $PBC$ and $ABC$ renormalized with nonvanishing external quasi-momenta, we would have gotten to the same regularization. The physical results are equivalent independently of our choice to define the theory either with zero or nonzero external quasi-momenta. 
\subsection{Normalization Conditions and critical exponents}
\par In the normalization conditions with null external quasi-momenta we will use the minimal set of diagrams with the three-loop level bare mass. The conventions utilized in Ref. \cite{Amit} will be useful to our purposes.  
\par The primitively divergent vertex parts are sufficient to renormalize all those which are renormalizable multiplicatively. A generic vertex part of $M$ "external legs" and $N$ insertions of composite operators with all external quasi-momenta equal to zero, are renormalized multiplicatively according to the rule ($(M,N) \neq (0,2)$):
\begin{equation}\label{23}
\Gamma_{R}^{(M,N)}(p_{n}, Q_{n'};0;0; g, m)= Z_{\phi}^{\frac{M}{2}} Z_{\phi^{2}}^{N} \Gamma^{(M,N)}(p_{n}, Q_{n'}; 0, 0; \lambda_{0}, \mu, \Lambda).
\end{equation} 
In the above equation $Z_{\phi}$ and $Z_{\phi^{2}}$ are the normalization functions, $\Lambda (\sim a^{-1}$, where $a$ is the "lattice constant") is the "cutoff" that shall be suppressed from now on: we are going to use dimensional regularization in the remainder of the discussion. 
\par Although not obvious, the cases under consideration are restricted to $n=1,...,M$, $n'=0,1,...,N$ and our condensed notation actually means {\it that all external null quasi-momenta are included in the argument of the vertex parts}, namely $\Gamma^{(M,N)}(p_{n}, Q_{n'}; 0, 0; \lambda_{0}, \mu, \Lambda) = \Gamma^{(M,N)}(p_{n}, Q_{n'}; \underset{M terms}{\underbrace{0,...,0}}, \underset{N terms}{\underbrace{0,...,0}}; \lambda_{0}, \mu, \Lambda)$. The notation
$\Gamma^{(M,0)}=\Gamma^{(M)}$ will be adopted henceforth. 
\par The normalization conditions for the primitively divergent bare vertex parts are defined by:
\begin{subequations}\label{24}
\begin{eqnarray}
&& \tilde{\Gamma}_{R}^{(2)}(k=0; 0; g, m) = m^{2},
\label{24a} \\
&& \frac{\partial\tilde{\Gamma}_{R}^{(2)}(k; 0; g, m )}{\partial k^{2}}\Big|_{k^{2}=0} \
= 1,\label{24b} \\
&& \Gamma_{R}^{(4)}(k_{l}=0;0; g, m ) \equiv \Gamma_{R}^{(4)}\Big|_{SP}  = 2\tilde{\sigma}g  , \label{24c}\\
&& \Gamma_{R}^{(2,1)}(k=0,Q=0; 0,0; g, m) \equiv \Gamma_{R}^{(2,1)}\Big|_{\overline{SP}} = 1 ,\label{24d}
\end{eqnarray}
\end{subequations}  
where $m$ is the renormalized mass and $g$ is the dimensionful coupling constant. It is convenient to write dimensionful coupling constants in terms of dimensionless ones. The dimensionful bare coupling constant is written  as 
$\lambda_{0}=\mu^{\epsilon} u_{0}$, whereas the renormalized one is defined by $g=\mu^{\epsilon} u$, where $u_{0}$ and $u$ 
are the dimensionless bare and renormalized coupling constants, respectively.
\par When written in terms of the dimensionless renormalized coupling constant, the normalization function $Z_{\phi}$ up to the desired order as a power series in $u$ has the form $Z_{\phi}= 1 + b_{1} u + b_{2} u^{2} + b_{3} u^{3}$. We prefer to define the function  $\bar{Z}_{\phi^{2}}= Z_{\phi}Z_{\phi}^{2}= 1 + c_{1} u + c_{2} u^{2}$, and write the bare dimensionless coupling constant as $u_{0}= u[1 + a_{1} u + a_{2} u^{2}]$.  
\par The diagrammatic expansion of the derivative of the bare vertex part $\Gamma^{(2)}$ is given by (see Appendix A)
\begin{subequations}\label{25}
\begin{eqnarray}
&& \frac{\partial\tilde{\Gamma}^{(2)}(k, 0; \lambda_{0}, \mu )}{\partial k^{2}}\Big|_{k^{2}=0} \
= 1 - B_{2} u_{0}^{2} + B_{3} u_{0}^{3},\label{25a}\\
&& B_{2}= -\frac{N+2}{144 \epsilon} \Bigl[1 - \frac{\epsilon}{4} + \epsilon \tilde{W}_{0}(0,\tilde{r}) \Bigr],\label{25b}\\
&& B_{3}=-\frac{(N+2)(N+8)}{648 \epsilon^{2}}\Bigl[1-\frac{\epsilon}{4} + \frac{3\epsilon}{2} \tilde{W}_{0}(0,\tilde{r})\Bigr].\label{25c}
\end{eqnarray}
\end{subequations}
\par Consider Eq. (\ref{24b}) above up to $O(u^{2})$. It implies that $b_{1}=0$ and $b_{2}=B_{2}$. We come back later to evaluate $b_{3}$. For the time being the simplest task is to determine $a_{1}$ and $a_{2}$ from Eq. (\ref{24c}). 
\par The diagrammatic expansion of the $\Gamma^{(4)}$ bare vertex part up to two-loops at zero external momenta reads
\begin{subequations}\label{26}
\begin{eqnarray}
&& \Gamma^{(4)}\Big|_{SP} = 2 \tilde{\sigma}u_{0}\mu^{\epsilon}[1 - A_{1} u_{0} 
+ (A_{2}^{(1)} + A_{2}^{(2)})u_{0}^{2}], \label{26a} \\
&& A_{1} = \frac{(N+8)}{6\epsilon}\Bigl[1-\frac{\epsilon}{2} + \epsilon \tilde{\zeta}_{0}(\tilde{r})\Bigr]  ,\label{26b}\\
&& A_{2}^{(1)} = \frac{(N^{2}+6N+20)}{36} \Bigl\{\frac{1}{\epsilon^{2}}(1- \epsilon 
+ 2\epsilon \tilde{\zeta}_{0}(\tilde{r}))\Bigr\}  ,\label{26c}\\
&& A_{2}^{(2)} = \frac{(5N+22)}{18\epsilon^{2}}\Bigl(1-\frac{\epsilon}{2} + 2\epsilon 
\tilde{\zeta}_{0}(\tilde{r})\Bigr)  , \label{26d}
\end{eqnarray}
\end{subequations}
Replacing this into its normalization condition, expanding $u_{0}(u)$ and using the value of $b_{2}$ just determined, the identification of terms of the same order in $u$ implies that the divergences in this bare vertex part are eliminated as long as  
\begin{subequations}\label{27}
\begin{eqnarray}
&& a_{1} = \frac{(N+8)}{6 \epsilon} \Bigl[1 - \frac{\epsilon}{2} + \epsilon \tilde{\zeta}_{0}(\tilde{r})\Bigr], \label{27a} 
\\
&& a_{2} =  \Bigl[\frac{(N+8)}{6 \epsilon}\Bigr]^{2} \Bigl[1 + 2\epsilon \tilde{\zeta}_{0}(\tilde{r})\Bigr] 
-\frac{(2N^{2}+41N+170)}{72\epsilon}.\label{27b}
\end{eqnarray}
\end{subequations}
\par We now compute $b_{3}$ from the diagrammatic expansion. We just need the values of $a_{1}, b_{2}$ and $B_{3}$ replaced in the normalization condition Eq.(\ref{24b}). After some algebra we find
\begin{equation}\label{28}
b_{3}= -\frac{(N+2)(N+8)}{1296\epsilon^{2}} \Bigl[1 -\frac{7 \epsilon}{4} + 3 \epsilon \tilde{\zeta}_{0}(\tilde{r})\Bigr].
\end{equation}
\par At this point we introduce the beta function  
\begin{eqnarray}\label{29}
&& \beta(u)=-\epsilon \Bigl(\frac{\partial\ln u_0}{\partial u}\Bigr)^{-1}= -\epsilon u
[1- a_{1} u + 2(a_{1}^{2} - a_{2}) u^{2} ],
\end{eqnarray}
whose ultraviolet fixed point of the coupling constant is defined by $\beta(u_{\infty})=0$. Exactly at this point the theory is scale invariant, and universal quantities like critical exponents can be computed. We then find
\begin{equation}\label{30}
u_{\infty}=\Bigl[\frac{6 \epsilon}{N+8}\Bigr] \Bigl\{1+\Bigl[\frac{(9N+42)}{(N+8)^{2}}+\frac{1}{2}
-\tilde{\zeta}_{0}(\tilde{r})\Bigr]\epsilon \Bigr\}.
\end{equation} 
\par The Wilson functions 
\begin{subequations}\label{31}
\begin{eqnarray}
&& \gamma_{\phi}(u)=\beta(u) \Bigl(\frac{\partial\ln Z_{\phi}}{\partial u} \Bigr)= -\epsilon u
[2 b_{2} u + (3 b_{3}- 2 b_{2} a_{1}) u^{2}], \label{31a}\\
&& \bar{\gamma}_{\phi^{2}}(u)= - \beta(u) \Bigl(\frac{\partial\ln \bar{Z}_{\phi^{2}}}{\partial u}\Bigr) = \epsilon u
[c_{1} + (2 c_{2} - c_{1}^{2} - a_{1} c_{1})u],\label{31b}
\end{eqnarray}
\end{subequations}
are related to the critical exponents $\eta$ and $\nu$. The first one is obtained up to three-loop order from the relation 
\begin{equation}\label{32}
\eta \equiv \gamma_{\phi}(u_{\infty})= \frac{(N+2)}{2(N+8)^{2}}\epsilon^{2} \Bigl\{1 + \epsilon \Biggl[\frac{6(3N+14)}{(N+8)^{2}} -\frac{1}{4}\Biggr]\Bigr\}.
\end{equation} 
\par The diagramatic expansion of the bare vertex part $\Gamma^{(2,1)}$ at zero external momenta is given by 
\begin{subequations}\label{33}
\begin{eqnarray}
&& \Gamma^{(2,1)}\Big|_{\overline{SP}}= 1 - C_{1} u_{0}
+ (C_{2}^{(1)} + C_{2}^{(2)})u_{0}^{2},\label{33a}\\
&& C_{1} = \frac{(N+8)}{6\epsilon}\Bigl[1-\frac{\epsilon}{2} + \epsilon \tilde{\zeta}_{0}(\tilde{r})\Bigr],\label{33b}\\
&& C_{2}^{(1)} = \frac{(N+2)^{2}}{36\epsilon^{2}}\Bigl[1- \epsilon  + 2 \epsilon \tilde{\zeta}_{0}(\tilde{r})\Bigr],\label{33c}\\
&& C_{2}^{(2)} = \frac{(N+2)}{12 \epsilon^{2}} \Biggl[1-\frac{\epsilon}{2} + 2 \epsilon \tilde{\zeta}_{0}(\tilde{r})\Biggr].\label{33d}
\end{eqnarray}
\end{subequations}
\par Following the same reasoning using the normalization condition for the composite field along with the results from Appendix A, $c_{1}$ and $c_{2}$ can be obtained and expressed in the form
\begin{subequations}\label{34}
\begin{eqnarray}
&& c_{1} = \frac{(N+2)}{6 \epsilon} \Bigl[1 - \frac{\epsilon}{2} + \epsilon \tilde{\zeta}_{0}(\tilde{r})\Bigr],\label{34a}\\
&& c_{2} = \frac{(N+2)(N+5)}{36 \epsilon^{2}} - \frac{(N+2)(2N+13)}{72\epsilon} + \frac{(N^{2}+7N+10)}{18 \epsilon}
\tilde{\zeta}_{0}(\tilde{r}).
\label{34b}
\end{eqnarray}
\end{subequations}
\par The exponent $\nu$ can be obtained using the identity $\nu^{-1}= 2 - \eta - \bar{\gamma}_{\phi^{2}}(u_{\infty})$. Therefore, up to two-loop order the critical index is given by
\begin{equation}\label{35}
\nu= \frac{1}{2} + \frac{(N+2)}{4(N+8)}\epsilon + \frac{(N+2)(N^{2}+23N+60)}{8(N+8)^{3}}\epsilon^{2}.
\end{equation} 
\par The results for the exponents are identical to those from the bulk using the massive framework. Plus, the 
agreement of the zero quasi-momenta renormalization approach to the primitively divergent vertex part for $NBC$ 
with the unified approach using nonzero external quasi-momenta in the renormalization algorithm for $NBC$ corroborates 
that the normal ordering procedure utilized in the finite size correction in the latter is consistent, since the physical results are the same in both methods even though the intermediary results have quite a different structure. 
\par Indeed, the correction disappears "miraculously" in the exponents. Moreover, in the limit $L \rightarrow 0$ the finite size correction diverges linearly in the limit $\tilde{r} \rightarrow \infty$ in the normal ordered expression of the unified approach as well as in the present approach with null quasi-momenta (which does not require any normal ordering prescription). This momentum space treatment is a reliable guide to unravel simple properties of finite critical systems. In order to have at hand a simple bird's-eye view of such systems, it is time to tackle the problem from the perspective of massless fields.    
\section{The renormalized massless theory}
\subsection{Consistency of the $\epsilon$-expansion in the massless theory}
\par The one-particle irreducible $(1PI$) formulation for finite size systems has no ambiguity, provided the critical system is kept away from the dimensional crossover region. In general, for finite values of $L$ the "scaling variable" 
$\frac{L}{\xi}$ in the massless regime ($\xi \rightarrow \infty$) is governed by the limit 
$\frac{L}{\xi} \rightarrow 0$. Nonetheless, this limit of the scaling variable is not sufficient to invalidate $\epsilon$-expansion results. 
\par Indeed, in Ref. \cite{NF2} it was recognized that $ABC$ and $DBC$ have the same critical exponents as the bulk system using arguments based purely on Green's functions results when the massless limit $\frac{L}{\xi} \rightarrow 0$ is taken. However, it was widely believed that in this limit the $\epsilon$-expansion results for the critical exponents from $PBC$ and $NBC$ were invalid. This belief proved incorrect. A direct update on the theory of finite size effects has begun with the consistent evaluation of critical exponents utilizing $1PI$ vertex functions with massless descriptions for $PBC$ and $NBC$ in Refs. \cite{BL,SBL}. We now examine the massless one-loop integrals of the four-point vertex part in an analogous treatment presented in the massive theory. The criterion for the breakdown of perturbation theory results should be unveiled within this picture.
\par  From the results in Appendix B for massless integrals we find   
\begin{subequations}\label{36}
\begin{eqnarray}
&& [I_{2}(k_{i};0; \tilde{\sigma}) + \tilde{I}_{2}(k=0;0;0; \tilde{\sigma})]\Big|_{SP} =  \frac{\kappa^{-\epsilon}}{\epsilon} \Bigl[1 + \frac{\epsilon}{2} + \epsilon \hat{\zeta}_{0}(\hat{r}) \Bigr], \label{36a}\\
&& \hat{\zeta}_{0}(\hat{r}) = \frac{1}{2} \Bigl[\int_{0}^{1} f_{\frac{1}{2}} (0, \hat{r}^{-1}\sqrt{x(1-x)}) + \pi \hat{r}\Bigr], \label{36b}\\
&& \hat{r}=\frac{\tilde{\sigma}}{\kappa},\label{36c}
\end{eqnarray}
\end{subequations}
where $SP$ is defined by $k_{i}.k_{j}= \frac{\kappa^{2}}{4}(4\delta_{ij} -1)$. The finite size correction 
$\hat{\zeta}_{0}(\hat{r})$ corresponds to the massless theory where $\kappa$ is the typical external momenta scale.
\par The first term of the finite size correction in the limit $L \rightarrow \infty$ was already shown to be zero for $PBC$ (analogous to $NBC$ here) and $ABC$ in Ref. \cite{BL} (see Eqs. (53)-(59) by setting $\tau =0$ and $\kappa^{2}$ therein). As $\hat{r} \rightarrow 0$ in this limit, $\hat{\zeta}_{0}(\hat{r}) (L \rightarrow \infty) \rightarrow 0$ consistent with the same limit in the massive formulation.
\par Consider the $L \rightarrow 0$ limit. Just as worked out in the massive theory,  at zero external quasi-momenta we do not need to devise any "normal ordering" and we obtain
\begin{equation}\label{37}
\underset{L \rightarrow 0}{lim} \;\;  \hat{\zeta}_{0}(\hat{r})  \rightarrow \; ln \hat{r}^{-1} + \pi  \hat{r},
\end{equation}
which is divergent, since $\hat{r}(L \rightarrow 0) \rightarrow \infty$, therefore invalidating $\epsilon$-expansion perturbative results. Except for a bigger coefficient in the dominant divergence (linear in $\hat{r}$) in comparison with the massive theory, the behaviors in both limits are entirely consistent with those coming from the massive theory. 
\par Surface effects coming from the linear divergence either in the massive or massless theories have the same origin: half of the correction comes from the diagonal integral $I_{2}$ and half come from the translation invariant symmetry breaking contribution coming from the integral $\tilde{I}_{2}$. The $PBC$ contribution associated with the integral $I_{2}$ has half the coefficient of the Neumann linear divergence in the limit $\hat{r} \rightarrow \infty$. The mechanism of dominance of $NBC$ in the bulk-surface transition (instead of the dominance of $PBC$ condition in that structural phase transition) is consistent with the analysis performed in the massive case. The coefficient of the linear divergence in the massless case is approximately three times bigger than its analogue in the massive theory, since the fluctuations are wilder at the transition temperature, justifying the enhancement of surface effects at the critical point as expected.  
\par Needless to say, the unified approach for $NBC$ when normal ordering of the finite size correction is taken into account shows exact agreement within the same physical limits in comparison with the simpler approach described in the present work. This proves that renormalization using either zero or nonzero external quasi-momenta for $NBC$ is immaterial, so long as the one-loop contribution of the four-point vertex part is concerned. Let us conclude the proof by performing the renormalization and evaluation of critical exponents at higher-loop orders.
\subsection{Renormalization and critical exponents}
\par By borrowing the arguments utilized in the massive setting, multiplicative renormalization is transliterated in the form ($(M,N) \neq (0,2)$):
\begin{equation}\label{38}
\Gamma_{R}^{(M,N)}(p_{n}, Q_{n'};0;0; g, m)= Z_{\phi}^{\frac{M}{2}} Z_{\phi^{2}}^{N} \Gamma^{(M,N)}(p_{n}, Q_{n'}; 0, 0; \lambda_{0}, \mu, \Lambda).
\end{equation} 
Dimensional regularization will be employed herafter and we forget about the cutoff $\Lambda$ henceforth. Just as before we adopt the convention $\Gamma^{(M,N)}(p_{n}, Q_{n'}; 0, 0; \lambda_{0}, \mu, \Lambda) = \Gamma^{(M,N)}(p_{n}, Q_{n'}; \underset{M terms}{\underbrace{0,...,0}}, \underset{N terms}{\underbrace{0,...,0}}; \lambda_{0}, \mu, \Lambda)$ and $\Gamma^{(M,0)}=\Gamma^{(M)}$. 
\par The normalization conditions for the primitively divergent bare vertex parts are defined by:
\begin{subequations}\label{39}
\begin{eqnarray}
&& \tilde{\Gamma}_{R}^{(2)}(k=0; 0; g, 0) = 0,
\label{39a} \\
&& \frac{\partial\tilde{\Gamma}_{R}^{(2)}(k; 0; g, 0 )}{\partial k^{2}}\Big|_{k^{2}=\kappa^{2}} \
= 1,\label{39b} \\
&& \Gamma_{R}^{(4)}(k_{l};0; g, 0 ) \equiv \Gamma_{R}^{(4)}\Big|_{SP}  = 2\tilde{\sigma}g  , \label{39c}\\
&& \Gamma_{R}^{(2,1)}(k_{1},k_{2},Q; 0,0; g, 0) \equiv \Gamma_{R}^{(2,1)}\Big|_{\overline{SP}} = 1 ,\label{39d}
\end{eqnarray}
\end{subequations}  
where $\overline{SP}$ is defined by $k_{i}^{2}=\frac{3}{4} \kappa^{2}$, $k_{1}.k_{2}= -\frac{1}{4}\kappa^{2}$ and 
$Q^{2}=\kappa^{2}$ (see \cite{Amit} for more details). Here, we write the dimensionful bare and renormalized coupling constant in terms of their dimensionless counterparts as $\lambda_{0}=\kappa^{\epsilon} u_{0}$ and $g=\kappa^{\epsilon} u$, respectively.
\par We write power series in $u$ of the renormalization functions, namely, $Z_{\phi}= 1 + b_{1} u + b_{2} u^{2} + b_{3} u^{3}$, $\bar{Z}_{\phi^{2}}= Z_{\phi}Z_{\phi}^{2}= 1 + c_{1} u + c_{2} u^{2}$, and the bare dimensionless coupling constant as $u_{0}= u[1 + a_{1} u + a_{2} u^{2}]$.  
\par  Using the results listed in Appendix B, we conclude that the diagrammatic expansion of the derivative of the bare vertex part $\Gamma^{(2)}$ reads
\begin{subequations}\label{40}
\begin{eqnarray}
&& \frac{\partial\tilde{\Gamma}^{(2)}(k, 0; \lambda_{0}, \mu )}{\partial k^{2}}\Big|_{k^{2}=\kappa^{2}} \
= 1 - B_{2} u_{0}^{2} + B_{3} u_{0}^{3},\label{40a}\\
&& B_{2}= -\frac{N+2}{144 \epsilon} \Bigl[1 + \frac{5 \epsilon}{4} - 2 \epsilon \hat{W}(\kappa,\hat{r}) \Bigr],\label{40b}\\
&& B_{3}=-\frac{(N+2)(N+8)}{648 \epsilon^{2}}\Bigl[1+ 2\epsilon - 3 \epsilon \hat{W}(\kappa,\hat{r})\Bigr].\label{40c}
\end{eqnarray}
\end{subequations}
\par Substitution of the last equation into Eq. (\ref{39b}) above up to $O(u^{2})$ leads to  $b_{1}=0$ and $b_{2}=-\frac{N+2}{144 \epsilon} \Bigl[1 + \frac{5 \epsilon}{4} - 2 \epsilon \hat{W}(\kappa,\hat{r}) \Bigr]$. We save the calculation of $b_{3}$ for later. From Eq. (\ref{39c}) we shall compute $a_{1}$ and $a_{2}$. 
\par At the symmetry point, the $\Gamma^{(4)}$ bare vertex part up to two-loops at zero external quasi-momenta is very similar to the massive vertex function, but with different coefficients. We have
\begin{subequations}\label{41}
\begin{eqnarray}
&& \Gamma^{(4)}\Big|_{SP} = 2 \tilde{\sigma}u_{0} \kappa^{\epsilon}[1 - A_{1} u_{0} 
+ (A_{2}^{(1)} + A_{2}^{(2)})u_{0}^{2}], \label{41a} \\
&& A_{1} = \frac{(N+8)}{6\epsilon}\Bigl[1+\frac{\epsilon}{2} + \epsilon \hat{\zeta}_{0}(\hat{r})\Bigr]  ,\label{41b}\\
&& A_{2}^{(1)} = \frac{(N^{2}+6N+20)}{36} \Bigl\{\frac{1}{\epsilon^{2}}(1 + \epsilon 
+ 2\epsilon \hat{\zeta}_{0}(\hat{r}))\Bigr\}  ,\label{41c}\\
&& A_{2}^{(2)} = \frac{(5N+22)}{18\epsilon^{2}}\Bigl(1+\frac{3 \epsilon}{2} + 2\epsilon 
\hat{\zeta}_{0}(\hat{r})\Bigr)  , \label{41d}
\end{eqnarray}
\end{subequations}
The renormalization algorithm follows the same flow of reasoning in the massless theory. Using Eq. (\ref{41}) into 
Eq. (\ref{39c}) yields directly  
\begin{subequations}\label{42}
\begin{eqnarray}
&& a_{1} = \frac{(N+8)}{6 \epsilon} \Bigl[1 + \frac{\epsilon}{2} + \epsilon \hat{\zeta}_{0}(\hat{r})\Bigr], \label{42a} 
\\
&& a_{2} =  \Bigl[\frac{(N+8)}{6 \epsilon}\Bigr]^{2} \Bigl[1 + 2\epsilon \hat{\zeta}_{0}(\hat{r})\Bigr] 
+\frac{(2N^{2}+23N+86)}{72\epsilon}.\label{42b}
\end{eqnarray}
\end{subequations}
\par A direct calculation of $b_{3}$ is now possible from the diagrammatic expansion as explained in the massive case. It is not difficult to show that
\begin{equation}\label{43}
b_{3}= -\frac{(N+2)(N+8)}{1296\epsilon^{2}} \Bigl[1 +\frac{5 \epsilon}{4} + 3 \epsilon \hat{\zeta}_{0}(\hat{r})\Bigr].
\end{equation}
\par The beta function  
\begin{eqnarray}\label{44}
&& \beta(u)=-\epsilon \Bigl(\frac{\partial\ln u_0}{\partial u}\Bigr)^{-1}= -\epsilon u
[1- a_{1} u + 2(a_{1}^{2} - a_{2}) u^{2} ],
\end{eqnarray}
now has the infrared fixed point in parameter space that is independent of the initial values of the bare parameters. The infrared fixed point of the coupling constant is defined by $\beta(u^{*})=0$. Recall that universal quantities like critical exponents can be computed after the determination of $u^{*}$, that is given by
\begin{equation}\label{45}
u^{*}=\Bigl[\frac{6 \epsilon}{N+8}\Bigr] \Bigl\{1+\Bigl[\frac{(9N+42)}{(N+8)^{2}}-\frac{1}{2}
-\hat{\zeta}_{0}(\hat{r})\Bigr]\epsilon \Bigr\}.
\end{equation} 
\par The Wilson functions 
\begin{subequations}\label{46}
\begin{eqnarray}
&& \gamma_{\phi}(u)=\beta(u) \Bigl(\frac{\partial\ln Z_{\phi}}{\partial u} \Bigr)= -\epsilon u
[2 b_{2} u + (3 b_{3}- 2 b_{2} a_{1}) u^{2}], \label{46a}\\
&& \bar{\gamma}_{\phi^{2}}(u)= - \beta(u) \Bigl(\frac{\partial\ln \bar{Z}_{\phi^{2}}}{\partial u}\Bigr) = \epsilon u
[c_{1} + (2 c_{2} - c_{1}^{2} - a_{1} c_{1})u],\label{46b}
\end{eqnarray}
\end{subequations}
at the infrared fixed point are related to the critical exponents $\eta$ and $\nu$. We obtain 
\begin{equation}\label{47}
\eta \equiv \gamma_{\phi}(u^{*})= \frac{(N+2)}{2(N+8)^{2}}\epsilon^{2} \Bigl\{1 + \epsilon \Biggl[\frac{6(3N+14)}{(N+8)^{2}} -\frac{1}{4}\Biggr]\Bigr\}.
\end{equation} 
\par The diagramatic expansion of the bare vertex part $\Gamma^{(2,1)}$ at zero external momenta can be written in the form
\begin{subequations}\label{48}
\begin{eqnarray}
&& \Gamma^{(2,1)}\Big|_{\overline{SP}}= 1 - C_{1} u_{0}
+ (C_{2}^{(1)} + C_{2}^{(2)})u_{0}^{2},\label{48a}\\
&& C_{1} = \frac{(N+8)}{6\epsilon}\Bigl[1+\frac{\epsilon}{2} + \epsilon \hat{\zeta}_{0}(\hat{r})\Bigr],\label{48b}\\
&& C_{2}^{(1)} = \frac{(N+2)^{2}}{36\epsilon^{2}}\Bigl[1 + \epsilon  + 2 \epsilon \hat{\zeta}_{0}(\hat{r})\Bigr],\label{48c}\\
&& C_{2}^{(2)} = \frac{(N+2)}{12 \epsilon^{2}} \Biggl[1 +\frac{3 \epsilon}{2} + 2 \epsilon \hat{\zeta}_{0}(\hat{r})\Biggr].\label{48d}
\end{eqnarray}
\end{subequations}
\par By employing the normalization condition for the composite field in conjumination with the results from Appendix B, $c_{1}$ and $c_{2}$ can be obtained. We find
\begin{subequations}\label{49}
\begin{eqnarray}
&& c_{1} = \frac{(N+2)}{6 \epsilon} \Bigl[1 + \frac{\epsilon}{2} + \epsilon \hat{\zeta}_{0}(\hat{r})\Bigr],\label{49a}\\
&& c_{2} = \frac{(N+2)(N+5)}{36 \epsilon^{2}} + \frac{(N+2)(2N+7)}{72\epsilon} + \frac{(N^{2}+7N+10)}{18 \epsilon} 
\hat{\zeta}_{0}(\hat{r}).
\label{49b}
\end{eqnarray}
\end{subequations}
\par The identity $\nu^{-1}= 2 - \eta - \bar{\gamma}_{\phi^{2}}(u^{*})$ leads to the result
\begin{equation}\label{50}
\nu= \frac{1}{2} + \frac{(N+2)}{4(N+8)}\epsilon + \frac{(N+2)(N^{2}+23N+60)}{8(N+8)^{3}}\epsilon^{2}.
\end{equation} 
\par The whole process confirms the universal character of the exponents, since they do not depend on the renormalization scheme either using massless or massive fields. Although the massless integrals are different and have a distinct typical scale (the external momenta scale $\kappa$, whereas in the massive case the proper scale is the bare mass $\mu$ at three-loop level) the critical exponents independ of the typical scale where the integrals are computed. The bottom line is: for finite $L$ and $\xi \rightarrow \infty$ at the critical point where the field theory becomes massless 
($\frac{L}{\xi} \rightarrow 0$), the physical results are well-defined and perturbation theory is really valid in the computation of critical indices for $ABC$, $PBC$, $DBC$ and $NBC$, in stark contrast to earlier beliefs regarding finite size effects.

\section{Conclusion}
\par In this paper we discussed the problem of renormalization of fields in a layered geometry in momentum space, confined in a finite size slab of thickness $L$ composed of $(d-1)$-dimensional parallel plates of infinite extent. We attacked the problem using null external quasi-momenta to begin with. The critical exponents obtained with Neuman boundary conditions applied in the limiting surfaces at $z=0,L$ coincide with those from the infinite system. Considering purely finite size effects here, the intermediate results strengthen our understanding why the dominance of $NBC$ in the description of the bulk-surface transition takes place.  
\par The present approach is simpler for $NBC$ in comparison with the recent unified approach proposed in Ref. \cite{SBL}: normal ordering of the finite size correction to the one-loop four point vertex part diagram is not necessary. The critical exponents for $NBC$ just obtained in this work are equal to those in the unified framework, to the $PBC$ and $ABC$ exponents from Ref. \cite{BL}: they are the same as those from the bulk system. Furthermore, the dimensional crossover criteria also coincide in both formalisms for $NBC$, providing a firmer ground for the dominance of the onset of surface effects purely out of volume fields in that case.   
\par Our treatment of massless fields in the the region $\frac{L}{\xi} \rightarrow 0$ for finite values of $L$ and away from the dimensional crossover region has a well-defined perturbation theory: not only the exponents are the same, but the dimensional crossover criteria are coincident either in the massless or in the massive setting, with minor modifications. The work just presented when collected with those from Refs. \cite{BL,SBL} represents a direct modern update in the theory of finite-systems: either massive or massless renormalizations to this problem in momentum space might improve our knowledge of more difficult effects appearing simultanenously with finite size ones. 
\par The study of critical amplitude ratios of several thermodynamic potentials require ingredients from the massive and massless framework {\it simultaneously}. With the modern approach to finite size effects on systems close to their critical points, we could update old results \cite{LNC,LSC} with the techniques described in our recent series of papers on finite-systems. Moreover, the computation of these amplitude ratios is lacking for the $ABC$ case, and it would be a nice exercise to see how finite size effects show up for $ABC$ in these ratios. Thin films \cite{F} could then be studied entirely using the modern techniques in momentum space. 
\par Finite-systems with Lifshitz bulk critical behavior could be studied within this new paradigm. This would allow the computation of universal quantities like critical exponents \cite{L1,L2,CL1,CL2} and amplitude ratios \cite{L3,L4,FL,SL}. Ferroelectric materials exhibit Lifshitz bulk behavior. Ferroeletric nanomaterials thin films might exhibit also ferromagnetism due to strain in the planes caused by the growth method and the substrate. The theoretical understanding of finite size effects in Lifshitz critical behavior could improve the understanding of more exotic boundary conditions, for instance, in ferroelectric semiconductor thin films \cite{EKM,Paz}. Another Lifshitz system is realized in multiferroic materials, but are more dificult to explain due to the ferrolectric and ferromagnetic coupling. This is a hot topic in the whole set of effects present in engineering nanomaterials \cite {Scott,subra}. We hope many other real physical finite-systems pertaining to other Lifshitz criticalities could have the starting point to explain their exotic properties by commencing with a similar simple description in the future along the same lines of the finite effects described so far.
\par After that, the study of surface effects by either introducing external surface fields in the Lagrangian density or by examining more carefully the dimensional crossover region in the present approach could shed new light in the mixing of finite size, interface and surface effects. Perhaps including both ingredients could give a clue on which effect is dominant. Therefore, this strategy has the potential to provide a better insight on the bulk-surface transition.
\par Beyond static aspects of critical phenomena, dynamic phase transitions in thin ferromagnetic films were also studied using Monte Carlo simulations for classical Heisenberg spin systems with certain classes of anisotropy and different boundary conditions for the order parameter \cite{JGH}. It might be interesting to study dynamical finite size states using field-theoretic methods applied to bulk systems \cite{Lovesey}. 
\par Another promising application for the field-theoretical techniques which have been developed for finite size 
systems concluded with the content of the present paper is the study of quantum finite-systems. Recent fundamentally simple ideas in determining a universal order parameter of quantum phase transitions utilizing finite size arguments \cite{SZB} have been put forth. In addition, studies on the modification of the topological phase transition mechanism induced by the finite size in interface-engineered $(Bi_{1-x}In_{x})Se_{3}$ thin films, using theoretical and experimental methods \cite{SSKBMO} indicates that very complex behaviors in the quantum character of finite size effects are waiting to be described with rigorous quantum field theory methods. It would be desirable to put together many theoretical tools to attack problems like these displaying simultaneously many non-trivial effects \cite{subra}.
\par The conclusion of the set of papers dealing with finite size effects employing only bulk fields in momentum space can pave the way to a complete description of more intrincate phenomena associated to the miniaturization intrinsic to nanodevices. We still have to envisage an efficient perturbation theory in the dimensional crossover region, perhaps using the smallness of $L$ dominating the poles in $\epsilon$. In this approach, we would have to consider all the integrals regular in $\epsilon$ since they depend on $L$, to determine their asymptotic (singular) $L$-dependence and neglect all dimensional poles. The logarithmic contributions in $L$ could be taken care of by postulating an inverse "finite size cutoff" $\mathcal{L}_{0}$ dominating the dimensional poles in $\epsilon$, or using another dimensional regularization where $ln \Bigl(\frac{L}{\mathcal{L}_{0}}\Bigr) \equiv ln \tilde{r}$ (massive theory) would be represented as dimensional poles in $\tilde{\epsilon}=1-d_{finite}$, whereas the linear divergence can be thought of as a single pole in $\tilde{\epsilon}$ in the same way quadratic ultraviolet divergences in standard field theory can be represented as a double pole in $\epsilon = 4-d$, at least in massive theories \cite{Kleinert}.
\par The next problem would be to devise a renormalization group for small 
$L$ or alternatively using $\mu$ or $\kappa$ as the running value since the $L$ dependence in the integral always come as the products $\kappa L$ or $\mu L$. The resulting treatment is not simple to guess in a glance and we live this and the above topics to be investigated in the future.
\section{Acknowledgments}
MVSS would like to thank financial support from CAPES, grant number 76640 and CNPq, grant number 141912/2012-0. JBSJ thanks  CNPq for financial support, grant number 142220/2007-8. MML acknowledges CNPq, grant number 232352/2014-3 for partial financial support.
\appendix
\section{Summary of massive integrals}
\par Since the integrals either massive or massless appearing in the present work were already discussed in the unified work from Ref. \cite{SBL} at nonnull external quasi-momenta, we shall simply list their solution in a suitable form to our purposes in the computation of the critical exponents. 
\par We list firstly the results of the integrals contributing to the four-point vertex part. At one-loop level the structure is very simple. In the two-loop graphs of the four-point vertex part we are interested at 
$O(\epsilon^{-1})$ and more singular terms. We note that the contributions of the four-point vertex part corresponding to the term $\overset{\infty}{\underset{j=-\infty}{\sum}}\tilde{I}_{2}^{2}(k;j,j;\tilde{\sigma}, \mu)$ is regular in 
$\epsilon$. It comes from the trivial two-loop graph Eq. (\ref{13}) and can be neglected. Similar remarks apply to the two-loop integral $\hat{I}_{4}(k,k';0,0,0;\tilde{\sigma},\mu)$ which is also regular in $\epsilon$ and are not going to 
be listed below. We should take advantage of the facts observed here for the four-point vertex part and substitute directly in the discussion of the vertex function $\Gamma^{(2,1)}$. 
\par The expression of every integral was previously derived in the main text. The results for the massive integrals corresponding to the four-point vertex part at zero external momenta and quasi-momenta are:
\begin{subequations}\label{A1}
\begin{eqnarray}
&& I_{2}(0;0;\tilde{\sigma}, \mu)= \mu^{-\epsilon} \Bigl[\frac{1}{\epsilon} \Bigl(1-\frac{\epsilon}{2}
+ \frac{\epsilon}{2} f_{\frac{1}{2}} (0, \tilde{r}^{-1})\Bigr)\Bigr] , \label{A1a}\\
&& \tilde{I}_{2}(0;0,0;\tilde{\sigma}, \mu) = \mu^{-\epsilon} \Bigl(\frac{\tilde{r}}{2}\Bigr),\label{A1b}\\
&& I_{4}(0,0;0,0;\tilde{\sigma}, \mu)= \mu^{-2\epsilon} \Bigl[\frac{1}{2 \epsilon^{2}} \Bigl(1-\frac{\epsilon}{2} 
+ \epsilon f_{\frac{1}{2}} (0, \tilde{r}^{-1})\Bigr)\Bigr] ,\label{A1c}\\
&& \tilde{I}_{4}(0;0,0,0;\tilde{\sigma}, \mu) = \mu^{-2\epsilon} \Bigl(\frac{\tilde{r}}{2 \epsilon}\Bigr).\label{A1d}
\end{eqnarray}
\end{subequations}
\par We now write down the results for the two-point vertex function integrals associated with two- and three-loop contributions. Let us emphasize that the three-loop integrals $\hat{I}_{5}(k;0,0,0; \tilde{\sigma}, \mu)$ and 
$\bar{I}_{5}(k;0,0,0;\tilde{\sigma},\mu)$ are regular in $\epsilon$ and their computation, therefore, will not concern us herein. Hence the relevant integrals of the two-point vertex parts which are going to be used in the normalization conditions as discussed in the text are actually their derivatives computed at zero external momenta. 
\par If we perform a rescaling in all momenta in the form $k' \rightarrow \frac{k}{\mu}$, the derivatives with respect with the new variable $k'$ is related to the derivative with respect to $k$ as $\frac{ \partial I_{3}(k;0;\tilde{\sigma}, \mu)}{ \partial k^{2}}=\frac{ \partial I_{3}(k';0;\tilde{r})}{\mu^{2} \partial k^{' 2}}$. Using this fact and expressing everything in terms of the scaled quantities, the two-loop integrals have the following solutions:
\begin{subequations}\label{A2}
\begin{eqnarray}
&& I_{3}^{'} (0;0;\tilde{\sigma}, \mu) \equiv \frac{ \partial I_{3}(k';0;\tilde{r})}{\mu^{2} \partial k^{' 2}}
\Bigl|_{k^{'2}=0} =  \mu^{-2\epsilon} \Bigl[-\frac{1}{8\epsilon}\Bigr] \Bigl(1-\frac{\epsilon}{4} 
+ \epsilon W_{0}(\tilde{r})\Bigr) , \label{A2a}\\
&& W_{0}(\tilde{r})= G_{0}(\tilde{r}) + H_{0}(\tilde{r}) - 4 F_{0}^{'}(\tilde{r}),\label{A2b}\\
&& G_{0}(\tilde{r})= -\frac{1}{2} - 2\int_{0}^{1} \int_{0}^{1}dx dy (1-y)ln\Bigl[(1-y)\tilde{r}^{-2} 
+ \frac{y\tilde{r}^{-2}}{x(1-x)}\Bigr],\label{A2c}\\ 
&& H_{0}(\tilde{r}) = 2 \int_{0}^{1} \int_{0}^{1}dx dy (1-y)f_{\frac{1}{2}}\left(0,\sqrt{(1-y)\tilde{r}^{-2} 
+ \frac{y\tilde{r}^{-2}}{x(1-x)}}\right),\label{A2d}\\
&& F_{0}^{'}(\tilde{r}) \equiv \frac{ \partial F_{0,1}(k';i;\tilde{r})}{ \partial k^{' 2}}
\Bigl|_{(k^{'2}=0, i=0)},\label{A2e}\\
&& F_{\alpha}(k',i;\tilde{r}) \equiv \tilde{r}^{-2 \alpha}\int_{0}^{1} dx f_{\frac{1}{2} + \alpha}(xi,\tilde{r}^{-1} 
\sqrt{x(1-x)(k^{'2} +\tilde{r}^{2}i^{2})+1}) ,\label{A2f}\\
&& F_{\alpha, \beta}(k',i,\tilde{r}) \equiv  \frac{1}{S_{d}} \tilde{r} \overset{\infty}{\underset{j=-\infty}{\sum}}
\int d^{d-1}q' \frac{F_{\alpha}(q'+k',j+i,\tilde{r})}{[q^{'2}+ \tilde{r}^{2}j^{2} + 1]^{\beta}},\label{A2g}\\
&& \tilde{I}_{3}^{'} (0;0;\tilde{\sigma}, \mu) \equiv \frac{\partial \tilde{I}_{3}(k';0,0;r)}{\mu^{2}  \partial k^{'2}}\Bigl|_{k^{'2}=0} = - \mu^{- 2 \epsilon}\Bigl[\frac{\tilde{r}}{4}
\int_{0}^{1}dx \int_{0}^{1} dy (1-y)\Bigl( 1-y 
+ \frac{y}{x(1-x)}\Bigr)^{-\frac{1}{2}}\nonumber\\
&&  -\frac{1}{2} \mathcal{F}_{0}^{'}(0;0,0;\tilde{r})\Bigr]. \label{A2h}
\end{eqnarray}
\end{subequations} 
In last equation we employed the definitions
\begin{subequations}\label{A3}
\begin{eqnarray}
&& \mathcal{F}_{\alpha,\beta}(k';0,0;\tilde{r}) \equiv \frac{1}{S_{d}} \tilde{r}  
\int d^{d-1}q' \frac{F_{\alpha}(q'+k';0;\tilde{r})}{[q^{' 2} + 1]^{\beta}},\\
&& \mathcal{F}_{\alpha}^{'}(0;0,0;\tilde{r})  \equiv \frac{\partial \mathcal{F}_{\alpha, 1}(k';0,0;\tilde{r})}{\partial k^{' 2}}\Bigl|_{k^{' 2}=0}.
\end{eqnarray}
\end{subequations}
\par The same comments about the scaling of the momenta are in order here. Therefore, one finds
\begin{subequations}\label{A4}
\begin{eqnarray}
&& I_{5}^{'}(0,0;0,0;\tilde{\sigma}, \mu)= \mu^{-3\epsilon} \Bigl[-\frac{1}{6 \epsilon^{2}}\Bigr] \Bigl(1-\frac{\epsilon}{4} + \frac{3 \epsilon}{2} W_{0}(\tilde{r})\Bigr) ,\label{A4a}\\
&& \tilde{I}_{5}^{'} (0;0,0,0;\tilde{\sigma}, \mu) = \mu^{-3 \epsilon} \Bigl(-\frac{1}{2 \epsilon}\Bigr)
\Bigl[\tilde{r} \int_{0}^{1}dx \int_{0}^{1} dy (1-y)\Bigl( 1-y 
+ \frac{y}{x(1-x)}\Bigr)^{-\frac{1}{2}}\nonumber\\
&&  - 2 \mathcal{F}_{0}^{'}(0;0,0;\tilde{r})\Bigr].\label{A4b}
\end{eqnarray}
\end{subequations} 
In order to make the connection with the content of the text we define the quantity 
\begin{equation}\label{A5}
\tilde{W}_{0}(0,\tilde{r}) = W_{0}(\tilde{r}) + 2 \tilde{r} \int_{0}^{1}dx \int_{0}^{1} dy (1-y)\Bigl( 1-y 
+ \frac{y}{x(1-x)}\Bigr)^{-\frac{1}{2}}  - 4 \mathcal{F}_{0}^{'}(0;0,0;\tilde{r}). 
\end{equation}
This amount is going to be used in the renormalization of the two-point vertex. However, it is demonstrated in the main text that it does not show up in the expression for the critical exponents.

\section{Formulae for massless integrals}
\par In the previous Appendix, most of the notation was already fixed. Except for minor modifications, which will be briefly highlighted here, we follow the same flow of reasoning. For instance, $\hat{r}=\frac{\tilde{\sigma}}{\kappa}$, where $\kappa$ is the typical external momenta scale. The external momenta $P$ is defined by $P=k_{1} + k_{2}$. Recall that in the total contribution of the four-point vertex function other permutations of the external momenta must be included.
\par The four-point vertex part contributions relevant to our purposes are represented by the one- and two-loop massless integrals computed at the symmetry point defined in the main text. At nonvanishing external momenta at the symmetry point and zero external quasi-momenta, their results are written down below directly, namely
\begin{subequations}\label{B1}
\begin{eqnarray}
&& I_{2}(P;0;\tilde{\sigma})\Bigl|_{SP}= \kappa^{-\epsilon} \Bigl[\frac{1}{\epsilon} \Bigl(1+\frac{\epsilon}{2}
+ \frac{\epsilon}{2} \int_{0}^{1} dx f_{\frac{1}{2}} (0, \hat{r}^{-1}[x(1-x)]^{\frac{1}{2}})\Bigr)\Bigr] , \label{B1a}\\
&& \tilde{I}_{2}(P;0,0;\tilde{\sigma})\Bigl|_{SP} = \kappa^{-\epsilon} \Bigl(\frac{ \pi \hat{r}}{2}\Bigr),\label{B1b}\\
&& I_{4}(P,k_{3};0,0;\tilde{\sigma})\Bigl|_{SP}= \kappa^{-2\epsilon} \Bigl[\frac{1}{2 \epsilon^{2}} \Bigl(1 +\frac{3 \epsilon}{2} 
+ \epsilon \int_{0}^{1} dx f_{\frac{1}{2}} (0, \hat{r}^{-1}[x(1-x)]^{\frac{1}{2}})\Bigr)\Bigr] ,\label{B1c}\\
&& \tilde{I}_{4}(P,k_{3};0,0,0;\tilde{\sigma})\Bigl|_{SP} = \kappa^{-2\epsilon} \Bigl(\frac{\hat{r}}{2 \epsilon}\Bigr).\label{B1d}
\end{eqnarray}
\end{subequations}
\par Let us focus on the two-point vertex part diagrams. Just as in the massive case, provided the system is away 
from the "dimensional crossover" region where the smallness due to $L$ dominates over the dimensional poles in 
$\epsilon$, we do not have to worry about regular contributions in $\epsilon$ in the the three-loop graph. Thus, we restrict ourselves to the presentation of the the results for $I_{3}, \tilde{I}_{3}, I_{5}$ and $\tilde{I}_{5}$. 
\par To begin with, we present the integrals of $I_{3}^{'}$ and $\tilde{I}_{3}^{'}$ that are given by:
\begin{subequations} \label{B2}
\begin{eqnarray}
&& \frac{\partial I_{3}(k;0,\tilde{\sigma})}{\partial k^{2}}\Bigl|_{k^{2}= \kappa^{2}} = - \kappa^{-2 \epsilon} \Bigl[\frac{1}{8 \epsilon}\Bigr] \Bigl(1 + \frac{5 \epsilon}{4} - 2 \epsilon W(\hat{r})\Bigr),\label{B2a}\\
&& W(\hat{r})= 2 \hat{F}_{0}^{'}(\hat{r}) - \hat{\bar{F}}_{0}(\hat{r}), \label{B2B}\\
&& \hat{F}_{\alpha}(k,i,\tilde{\sigma})= \frac{\tilde{\sigma}^{-2 \alpha}}{S_{d}} \int_{0}^{1} dx 
f_{\frac{1}{2} + \alpha}\Bigl(0, \Bigl[\Bigl(\frac{k^{2}}{\tilde{\sigma}^{2}} + i^{2}\Bigr)x(1-x)\Bigr]^{\frac{1}{2}}\Bigr), \label{B2c}\\
&& \hat{F}_{\alpha,\beta}(k,i=0,\tilde{\sigma}) = \frac{\tilde{\sigma}}{S_{d}}\overset{\infty}{\underset{j=-\infty}{\sum}}
\int d^{d-1}q \frac{\hat{F}_{\alpha}(q+k,j,\tilde{\sigma})}{[q^{2}+ \tilde{\sigma}^{2}j^{2}]^{\beta}},\label{B2d}\\
&& \hat{F}_{0}^{'}(\hat{r}) = \frac{\partial \hat{F}_{0,1}(k,i=0,\tilde{\sigma})}{\partial k^{2}}\Bigl|_{k^{2} = \kappa^{2}},\label{B2e}\\
&& \hat{\bar{F}}_{0}(\hat{r})=  \int_{0}^{1} dx (1-x) f_{\frac{1}{2}}(0, \sqrt{x(1-x)}\hat{r}^{-1}), \label{B2f}\\ 
&& \frac{\partial I_{3}(k;0,\tilde{\sigma})}{\partial k^{2}}\Bigl|_{k^{2}= \kappa^{2}}= - \kappa^{-2 \epsilon} \Bigl[\frac{\hat{r}}{4} \hat{H}_{0} -\frac{\kappa^{2 \epsilon}}{2}
\hat{\mathcal{F}}_{0}^{'}(\hat{r})\Bigr],\label{B2g}\\
&& \hat{H}_{0}= \int_{0}^{1} dx (1-x)[x(1-x)]^{-\frac{1}{2}}, \label{B2h}\\
&& \hat{\mathcal{F}}_{\alpha,\beta}(k;i,j;\tilde{\sigma})= \frac{\tilde{\sigma}}{S_{d}} \int \frac{d^{d-1}q 
\hat{F}_{\alpha}(q+k,j,\tilde{\sigma})}{[q^{2}+ \tilde{\sigma}^{2}i^{2}]^{\beta}}, \label{B2i}\\
&& \hat{\mathcal{F}}_{0}^{'}(\hat{r}) \equiv \frac{\partial \hat{\mathcal{F}}_{0,1}^{(\tau)}(k,i=0,j=0;\tilde{\sigma})}
{\partial k^{2}}\big|_{k^{2}=\kappa^{2}}. \label{B2j}
\end{eqnarray}
\end{subequations}
\par By defining $\hat{\mathcal{H}}_{0}(\kappa, \hat{r}) = \kappa^{2 \epsilon} \hat{\mathcal{F}}_{0}^{'}(\hat{r})$, the combination that will be helpful in our discussion in the main text regarding the renormalization conditions theme is defined by
\begin{equation}\label{B3}
\hat{W}(\kappa, \hat{r})= W(\hat{r}) - \hat{r} \hat{H}_{0} + 2 \hat{\mathcal{H}}_{0}(\kappa, \hat{r}).
\end{equation}
\par With all these definitions, it is not difficult to evaluate $I_{5}^{'}$ and $\tilde{I}_{5}^{'}$. The results are:
\begin{subequations} \label{B4}
\begin{eqnarray}
&& I_{5}^{'}(k,0,\tilde{\sigma})\Bigl|_{k^{2}=\kappa^{2}} = -\frac{\kappa^{-3 \epsilon}}{6 \epsilon^{2}}\Bigl[1+2 \epsilon -3 \epsilon W(\hat{r})\Bigr], \label{B4a}\\
&& \tilde{I}_{5}^{'}(k,0,\tilde{\sigma})\Bigl|_{k^{2}=\kappa^{2}} =  -\frac{\kappa^{-3 \epsilon}}{\epsilon}
\Bigl[\frac{\hat{r}}{2} \hat{H}_{0} - \hat{\mathcal{H}}_{0}\Bigr].\label{B4b}
\end{eqnarray}
\end{subequations}

\end{document}